# Water Production of Interstellar Comet 3I/ATLAS from SOHO/SWAN Observations after Perihelion

M.R. Combi[1], T. Mäkinen[2], J.-L. Bertaux[3], E. Quémerais[3], S. Ferron[4], R. Lallement[5] and W. Schmidt[2]

[1]Dept. of Climate and Space Sciences and Engineering
University of Michigan
2455 Hayward Street
Ann Arbor, MI 48109-2143
USA
*Corresponding author: mcombi@umich.edu

[2]Finnish Meteorological Institute, Box 503
SF-00101 Helsinki, Finland

[3]LATMOS/IPSL
UVSQ Université Paris-Saclay, Sorbonne Université, CNRS,
Guyancourt, France

[4]ACRI-st, Sophia-Antipolis, France

[5]LIRA, Observatoire de Paris, Meudon, France

## ABSTRACT

The Solar Wind ANisotropies (SWAN) all-sky hydrogen Lyman-alpha camera on the Solar and Heliosphere Observatory (SOHO) observed the hydrogen coma of interstellar comet 3I/ATLAS, also called C/2025 N1 (ATLAS), beginning on November 6, 2025, 9 days after perihelion. Water production rates were calculated from each image of 3I/ATLAS using the methodology of Mäkinen & Combi (2005, Icarus 177, 217) and fluorescence rates, g-factors, calculated using the daily solar Lyman-alpha fluxes from the LASP database (https://lasp.colorado.edu/lisird/data) corrected for solar rotation and for the comet's heliocentric velocity. The method has been used for over 90 comet apparitions (Combi 2022; Combi et al. 2019). A water production rate of 3.17 x $10^{29}$ $s^{-1}$ was found on November 6 when the comet was at a heliocentric distance of 1.40 au and at a sufficient solar elongation angle. It decreased over time after that, down to 1-2 x $10^{28}$ $s^{-1}$ around 40 days post-perihelion (December 9).

## Introduction

The third known object to originate outside of our solar system and second comet, 3I/ATLAS = C/2025 N1 (ATLAS), was discovered with the Asteroid Terrestrial-impact Last Alert System (ATLAS) 0.5-meter telescope in Rio Hurtado, Chile, on 1 July 2025 (Denneau et al. 2025). It was almost immediately known to have a surrounding coma (Minev et al. 2025), and pre-discovery observations (e.g., Martinez-Palomera et al. 2025) confirmed its interstellar origin with a hyperbolic orbit with eccentricity of 6 (Minor Planets Center). Its orbital inclination is ~175 degrees making its orbit retrograde and only ~5 degrees out of the plane of the solar system (ecliptic). Because this comet comes from essentially infinity, its orbital inclination could have any value with equal probability.

The first confirmed interstellar object, 1I/(2017 U1) ’Oumuamua, did not have a detected gaseous or solar-continuum scattered dust coma surrounding it. Comet 2I/Borisov, the first interstellar comet to be discovered, on the other hand, had a well-developed coma with CN (Fitzsimmons et al. 2019), a common cometary coma gas species, and dust and water (McKay et al. 2020; Xing et al. 2020). It was found to be carbon chain depleted, similar to the class of carbon-chain depleted comets of solar system origin (Kareta et al. 2020).

Early observations of 3I/ATLAS indicated that the comet was also carbon chain depleted but showed a strong CN emission (Salazar Manzono et al. 2025). Observations of 3I/ATLAS after perihelion on November 29 (Jehin et al. 2025a) show strong $C_2$ and $C_3$ emissions with a more typical $C_2$/CN ratio with log(Q-$C_2$/Q-CN) = +0.04.

Measurements of the gases and byproducts giving clues to the main volatile constituents then followed. Xing et al. (2025) observed OH in 3I with the Swift satellite on 31 July – 1 August, 2025, when the comet was at a heliocentric distance of 2.9 au. These indicated a total water production rate of (0.74±0.50) x $10^{27}$ $s^{-1}$. JWST observations on August 6 (Cordiner et al. 2025) measured a lower water production rate of (2.23±0.08) x $10^{26}$ $s^{-1}$, whereas later observations with Swift on August 18-20 found again a much larger water production rate of (1.26±0.35) x $10^{27}$ $s^{-1}$. Xing et al. suggested that the difference might be due to an extended source of icy particles released and moving away from the nucleus that would contribute to larger apparent water production rates in the large field of view of Swift compared with the smaller aperture of JWST. The JWST observations of Cordiner et al. (2025) also provided a $CO_2$ production rate of (1.70±0.01) x $10^{27}$ $s^{-1}$ and of CO of 3.7±0.02) x $10^{26}$ $s^{-1}$, both larger than that of $H_2O$. At a heliocentric distance of 3.32 au, this is not too surprising, as the post-perihelion measurements at > 3 au of the relative production rates of $CO_2$ and CO were also larger or comparable to that of $H_2O$ in the Rosetta target comet 67P/Churyumov-Gerasimenko (Combi et al. 2020).

Hutsemekers et al. (2025) reported log OH production rates from ground-based echelle spectroscopic observations of 3I/ATLAS with the VLT on 8 August, 3-4 September, 10 September, and 12 September of 26.37 ± 0.04, 26.59 ± 0.02, 27.05 ± 0.13 and 27.15 ± 0.10. These give water production rates of approximately 2.8, 4.6, 13.4, and 16.8 in $10^{26}$ $s^{-1}$.

Crovisier et al. (2025) reported the results of 1667-MHz observations OH 18-cm line with the Nançay radio telescope during the period of 13 – 19 October 2025. They found an OH production rate of (5.7±0.6) x $10^{28}$ $s^{-1}$ which scales to an approximate water production rate of 6.7 x $10^{28}$ $s^{-1}$. This amounts an increase in water production by a factor of 60 over 30 days from the earlier observations.

Jehin et al. (2025a) found an OH production rate on November 29 of 1.67 ± 0.27 x $10^{28}$ $s^{-1}$ corresponding to a water production rate of about 1.98 x $10^{28}$ $s^{-1}$ assuming an OH/$H_2O$ yield of 0.84. Jehin et al. (2025b) reported a smaller OH production rate of (6.93 ± 1.13) x $10^{27}$ $s^{-1}$ corresponding to a water production rate of 8.4 x $10^{27}$ $s^{-1}$ on December 7. These are notably smaller than those from Crovisier et al. (2025).

**$H_2O$ Rates from SWAN observations of H Ly-alpha**

The Solar Wind Anisotropies (SWAN) all-sky Lyman-alpha camera on the Solar and Heliosphere Observatory (SOHO) satellite located in a halo orbit around the L1 Lagrange point between the Earth and Sun (Bertaux et al. 1995) observed the hydrogen Lyman-alpha coma of 3I/ATLAS beginning 9 days after perihelion, once it emerged from the solar obstruction area of SWAN (blind zone in Figure 2). It was not detectable when it was bright enough around and before perihelion, as it was located in the solar obstruction area. Because solar UV photodissocation of water produces a net of 2 H atoms per water molecule with OH as an intermediary for half, the water production rate of the comet can be calculated from an analysis of the H coma with a well-established procedure (Mäkinen & Combi 2005). To date SWAN has observed over 90 comet apparitions, including multiple apparitions each of several short period comets over the past 30 years and many long-period Oort Cloud comets (Combi et al. 2019; Combi 2022). A special operation procedure was implemented in the SWAN data operation procedure to move a regularly occurring artifact that shows itself on one side next to the spacecraft obstruction in the full sky images to the opposite side. This enabled obtaining an extra weeks' worth of results to be obtained for 3I.

Table 1 gives the observational circumstances and the water production rates as a function of time in days from perihelion on 29.48 October 2025. The uncertainty given is the formal uncertainty associated with the fit of the coma model and interplanetary background to the image of the comet which is rather small. It does not account for uncertainties in the instrument calibration, the solar Lyman-alpha flux calibration, or the photochemical lifetimes and other parameters in the coma model. There are also contributions to uncertainty from unaccounted for field stars which are numerous with SWAN's 1-degree pixels. We have estimated all those could amount to up to 30%. Figure 1 shows the water production rates plotted also as a function of time from perihelion. Figure 2 shows a representative sample of a SWAN image of 3I/ATLAS from 12 November 2025. For very bright comets observed with SWAN that have higher signal-to-noise and cover much more of the sky the shape of the coma influenced by radiation pressure is more evident (Combi et al. 2000). This is clearly not the case for 3I/ATLAS or many fainter comets.

The first few values of the production rate agree extremely well with those determined from OH spectra measurements made over a very wide field of view encompassing essentially all the OH in the coma out to several hundred thousand km with the MAVEN spacecraft UVIS instrument (M. Stevens and J. Deighan, private communication) over the same time period. A description of the MAVEN UVIS instrument is given by McClintock et al. (2014). Observations by MAVEN UVIS of comet C/2013 A1 (Siding Spring) were described by Crismani et al. (2015) and Mayyasi et al. (2020). Those measurements were fairly constant between 2.2 and 3.2 x $10^{29}$ $s^{-1}$ going back to 6 days before perihelion. Because they include the whole coma, they are only dependent on g-factor and the lifetime of OH radicals. Both MAVEN and SWAN therefore include all water production from the nucleus and any extended sources if present.

Ground-based observations of OH with the Trappist telescope (Jehin et al. 2025a&b, and private communication) from late November to early December are consistent with water production rates that are 7 times smaller than SWAN on 29 November to 2 times smaller on December 7. Similar discrepancies that seem to depend on the size of the field-of-view (fov) of the observation were noted by Xing et al. (2025) who had observations of OH from the Swift satellite compared with those very small fov observations of H2O from Cordiner et al. (2025) using JWST. About this apparent discrepancy, Xing et al. raised a possible role of an extended source of water from released solid ice particles that subsequently sublimate at larger distances from the nucleus. This would also be consistent with the possible detection of water ice features by Lisse et al. (2025).

Long period Oort Cloud comet C/2006 P1 (Garradd) showed strong evidence of an important extended water source from icy grains (Combi et al. 2013; Bodewits et al. 2014). Similarly, Jupiter Family comet 103P/Hartley 2 has been shown directly and indirectly to have a substantial extended source icy grains (A'Hearn et al., 2011; Hermalyn et al., 2013; Kelley et al., 2013; Protopapa et al., 2014; Shou et al. 2025; Feaga & Sunshine 2025) which might explain so-called hyperactive comets generally speaking (Khan et al. 2023). If an extended source from icy grains is responsible for the larger water production rates obtained with larger fields of view then the contribution from those icy grains decreases substantially between early November and early December.

**Discussion**

From the SWAN data (9 to 41 days post-perihelion, figure 1) it can be computed that a total of 13.5 millions metric tons of $H_2O$ has been released from the comet during this period. The total active area of 3I/ATLAS ranged from a high of 193.7 $km^2$ down to 9.9 $km^2$ using the model of Cowan & A'Hearn (1979). This decrease would be consistent with the relative decrease in the contribution of an extended source of icy grains over this period. If water production came exclusively from a simple rotating nucleus, then the minimum radius would correspond to only 0.88 km if the surface were 100% active, only slightly larger than Jupiter Family Comets 46P/Wirtanen and 103P/Hartley 2 and similar to that of Oort Cloud comet 1996 B2 (Hyakutake); see e.g., Combi et al. (2005). See also the extensive table in the paper by Combi et al. (2019). Comets that seem to be nearly 100% (or more) active appear to have an extensive contribution to water production from an extended source of released icy particles (Khan et al. 2023).

$H_2O$ production rates are an essential reference for all other gas production rates for the chemical and isotopic composition of volatile compounds in comets. The D/H ratio in water is of particular interest, as it constrains the deuterium abundance in primitive ice grains. Comet formation models indicate that, in the solar system, the D/H ratio of primitive ice was about twice larger than the average D/H ratio in comets (e.g. Kavelaars et al. 2011). If we make the reasonable assumption that the formation process of comets is similar in other stellar systems, then the D/H ratio for interstellar comets such as 3I/ATLAS would also indicate an about twice larger D/H ratio in its pre-stellar nebula. It should then provide valuable information about the origin of the observed variability of the D/H ratio in the ISM gas. As a matter of fact, one debated interpretation is that all D atoms missing in the gas have accumulated into icy grains. In this case, up to four orders of magnitude of deuteration are predicted to occur in ices of very dense clouds, based on the D/H depletion measured in interstellar gas along several lines of sight (Draine, 2004, Linsky et al. 2006, Lallement et al., 2008, Friedman et al., 2023). Alternately, in view of the suggested release of icy particles from the surface of the nucleus, it cannot be totally excluded that what is seen in the coma comes from a layer of interstellar material that would have been accreted during one or several crossings of the nucleus through very dense and cold interstellar clouds, during its long Galactic journey.

Finally, we want to emphasize that the larger water production rates from observations over larger fields of view means that calculations of D/H ratios in 3I using different instruments for HDO and $H_2O$ may need to be interpreted more carefully.

**Acknowledgements:** SOHO is an international mission between ESA and NASA. M. Combi acknowledges support from NASA grant 80NSSC23K0030 from the Solar System Observations Program. T.T. Mäkinen and W.Schmidt were supported by the Finnish Meteorological Institute (FMI). J.-L. Bertaux and E. Quémerais acknowledge support from CNRS and CNES. We obtained orbital elements from the JPL Horizons web site (http://ssd.jpl.nasa.gov/horizons.cgi). For classification of the dynamical ages of the comets we used the Minor Planets Center web site https://www.minorplanetcenter.net/db_search tool. The composite solar Lyα data were taken from the LASP web site at the University of Colorado (https://lasp.colorado.edu/lisird/data). We acknowledge the personnel who have been keeping SOHO and SWAN operational for 30 years. We thank the MAVEN UVIS team for sharing the preliminary results of their observations before publication. We especially thank the reviewer, Jacques Crovisier, whose review helped to greatly improve this paper.

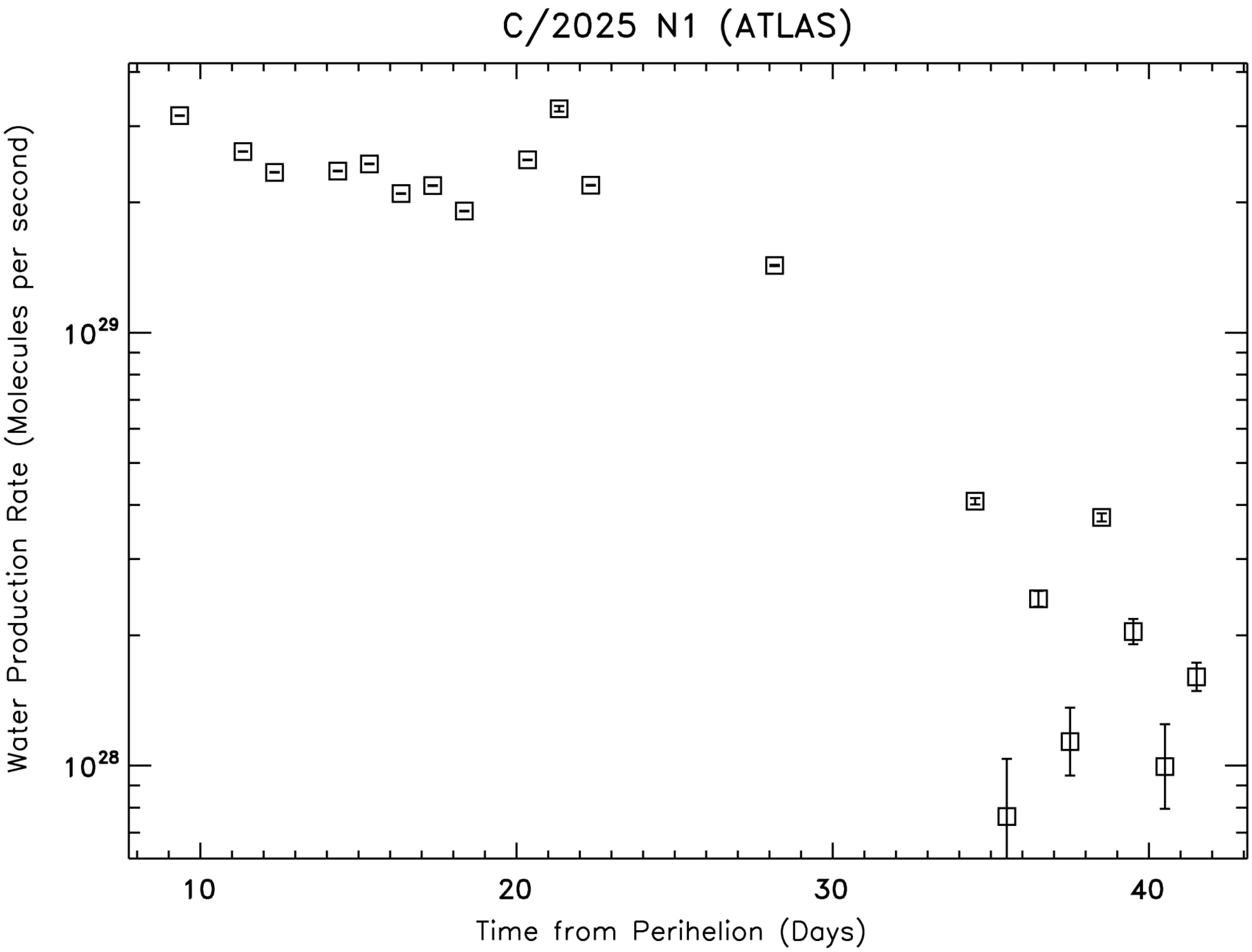

Figure 1. Water production rate of comet 3I/ATLAS as a function of time from perihelion (29.48 October 2025). The open squares give the water production rate in $s^{-1}$ from single images. The error bars give the 1-σ formal random fitting errors for each value. There is a potential ~30% uncertainty from the model parameters and calibration not included in the plotted error bars.

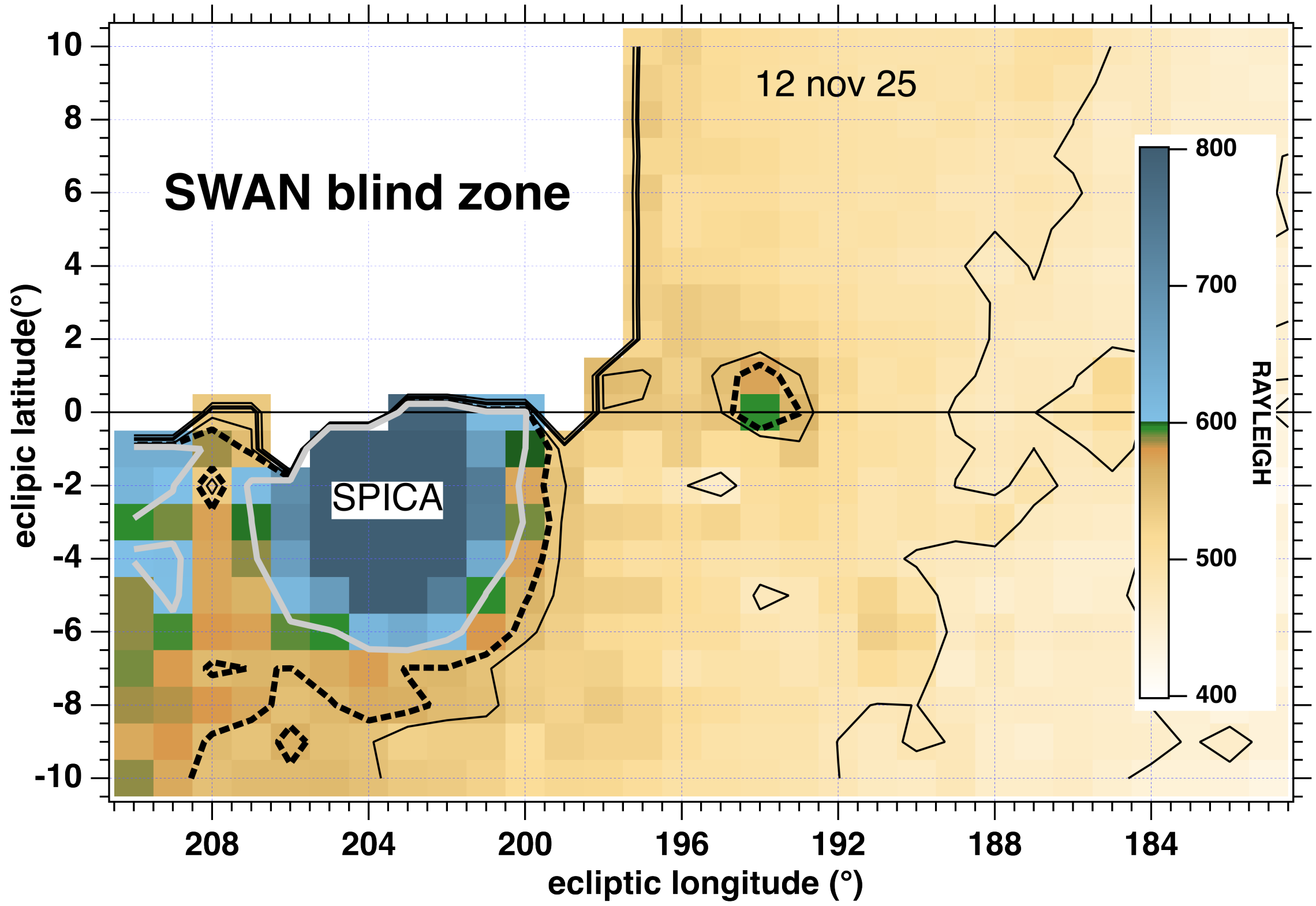

Figure 2. Lyman-α intensity map in the region of the comet 3I/ATLAS (Rayleigh, R). The hydrogen cloud is clearly visible around 195° ecliptic longitude, near 0° latitude. The resolution is 1° per pixel. It is a small portion of the full-sky map recorded on 12 November 2025. There is a blind zone which cannot be observed with SWAN, too near the sun (outside the map, on the left). The very bright UV star SPICA is just below the blind zone. The intensity is color coded, and some particular contour isophotes are drawn: thick dashed black line is at 560 R, thick grey line is at 600, the other thin lines are respectively at 545 R (one around the comet), and 400, 450, 480. The 545 R isophote around the comet (at distance to Earth 2.13 AU) has an approximate size of 16 x $10^6$ km in longitude by 13 x $10^6$ km in latitude.

Table 1

SOHO/SWAN Observations of 3I/ATLAS and Water Production Rates

| ΔT (Days) | R (au) | Δ (au) | g ($s^{-1}$) | Q ($10^{28}$ $s^{-1}$) | δQ ($10^{28}$ $s^{-1}$) |
|---|---|---|---|---|---|
| **9.344** | **1.401** | **2.187** | **0.002153** | **31.72** | **0.04** |
| **11.345** | **1.420** | **2.161** | **0.002348** | **23.10** | **0.03** |
| **12.345** | **1.431** | **2.149** | **0.002266** | **23.46** | **0.04** |
| **14.345** | **1.456** | **2.123** | **0.002202** | **23.62** | **0.04** |
| **15.345** | **1.470** | **2.110** | **0.002181** | **24.54** | **0.04** |
| **16.345** | **1.484** | **2.097** | **0.002197** | **20.96** | **0.04** |
| **17.346** | **1.499** | **2.084** | **0.002179** | **21.86** | **0.04** |
| **18.346** | **1.514** | **2.071** | **0.002194** | **19.09** | **0.04** |
| **20.346** | **1.548** | **2.045** | **0.002270** | **25.08** | **0.04** |
| **21.346** | **1.565** | **2.032** | **0.002314** | **32.89** | **0.49** |
| **22.346** | **1.584** | **2.019** | **0.002317** | **21.90** | **0.04** |
| **28.162** | **1.699** | **1.946** | **0.002407** | **14.30** | **0.04** |
| **34.503** | **1.841** | **1.874** | **0.002750** | **4.08** | **0.07** |
| **35.504** | **1.865** | **1.864** | **0.002785** | **0.76** | **0.27** |
| **36.504** | **1.888** | **1.854** | **0.002799** | **2.43** | **0.10** |
| **37.504** | **1.912** | **1.845** | **0.002703** | **1.14** | **0.22** |
| **38.504** | **1.937** | **1.836** | **0.002649** | **3.68** | **0.08** |
| **39.504** | **1.961** | **1.827** | **0.002616** | **2.04** | **0.14** |
| **40.505** | **1.986** | **1.819** | **0.002626** | **1.00** | **0.25** |
| **41.505** | **2.011** | **1.811** | **0.002636** | **1.60** | **0.13** |

Notes. ΔT (Days from Perihelion 29.48 October 2025)

r: Heliocentric distance (au)

Δ: Comet-SOHO distance (au)

g: Solar Lyman-α g-factor (photons $s^{-1}$) at 1 au

Q: Water production rates for each image ($s^{-1}$)

δQ: internal 1-sigma uncertainties